\documentclass[twocolumn,superscriptaddress,floatfix,prb]{revtex4-1}
\usepackage{graphicx,amsfonts,amssymb,amsmath,hyperref,hypcap,enumerate}
\usepackage{color}
\usepackage{cleveref}

\newcommand{\qq}{\boldsymbol{q}}

\newcommand{\rr}{\boldsymbol{r}}

\newcommand{\kk}{\boldsymbol{k}}

\newcommand{\bb}{\boldsymbol{b}}
\newcommand{\jj}{\boldsymbol{j}}

\newcommand{\bv}{\boldsymbol{v}}

\begin{document}
\title{Topological chiral superconductivity with spontaneous vortices and supercurrent in twisted bilayer graphene}

\author{Fengcheng Wu}
\email{wufcheng@umd.edu}
\affiliation{Condensed Matter Theory Center and Joint Quantum Institute, Department of Physics, University of Maryland, College Park, Maryland 20742, USA}
\affiliation{Materials Science Division, Argonne National Laboratory, Argonne, Illinois 60439, USA}

\date{\today}

\begin{abstract}
We study $d$-wave superconductivity in twisted bilayer graphene and reveal  phenomena that arise due to the moir\'e superlattice. In the $d$-wave pairing, the relative motion (RM) of two electrons in a Cooper pair can have either $d+id$ or $d-id$ symmetry with opposite angular momenta. Due to the enlarged moir\'e superlattice, the center-of-mass motion (COMM) can also carry a finite angular momentum  while preserving the moir\'e periodicity. By matching the total angular momentum, which has contributions from both the RM and the COMM, Cooper pairs with $d+id$ and $d-id$ RMs are intrinsically coupled in a way such that the COMM associated with one of the RMs has a spontaneous vortex-antivortex lattice configuration. Another  phenomenon is that the chiral $d$-wave state carries spontaneous bulk circulating supercurrent. The chiral $d$-wave superconductors are gapped and also topological as characterized by an integer Chern number. Nematic $d$-wave superconductors, which could be stabilized, for example, by uniaxial strain, are gapless with point nodes. 
\end{abstract}

\maketitle

\section{Introduction}
The twist angle in van der Waals bilayers has emerged as a new tuning knob to control electronic properties.\cite{TBL2007,Hunt2013,Dean2013,Wang2015,Kim2017,Chen2018} Theory has predicted that the Dirac velocity of twisted bilayer graphene (TBLG) vanishes at a set of magic twist angles \cite{Bistritzer2011}, near which the low-energy moir\'e bands are extremely flat and electron interaction effects are therefore magnified.
Correlated insulating states and superconductivity have recently been experimentally observed in TBLG near the largest magic angle ($\sim 1^\circ$).\cite{Cao2018Super,Cao2018Magnetic} These discoveries have generated  great interest in moir\'e pattern physics.\cite{Wu2018Hubbard, YHZhang2018,Jung2018, wu2018layer} A recent  experiment demonstrated that superconductivity in TBLG can be further tuned by pressure.\cite{Dean2018Tuning} In theory, various aspects of TBLG are being actively studied, including single-particle band structure theory  \cite{Fu2018,Koshino2018, Kang2018,Senthil2018, Bernevig2018Topology,Po2018faithful,Ahn2018failure,hejazi2018multiple,liu2018complete,Carr2018, chittari2018pressure, Vishwanath2018origin, Lado2018}, many-body theory on the low-temperature superconducting and correlated insulating states \cite{Senthil2018, Balents2018,roy2018unconventional, Dodaro12018, Padhi2018, Guo2018, huang2018AF, Liu2018chiral,Fidrysiak2018, Heikkila2018, rademaker2018charge, Kennes2018strong, Isobe2018, You2018, wu2018emergent,PALee2018, Wu2018phonon, guinea2018electrostatic,Thomson2018,gonzalez2018kohn, Lin2018, Lian2018twisted, sherkunov2018novel, Venderbos2018, Kozii2018, choi2018electron,wu2018coupled}, and also transport theory  in the high-temperature metallic regime \cite{Wu2019rho}.

In this work, we study $d$-wave superconductivity in TBLG, and reveal  phenomena that arise due to the enlarged moir\'e superlattices. The $d$ wave has been proposed to be a candidate pairing symmetry for TBLG in Coulomb repulsion mechanism \cite{Balents2018,Liu2018chiral,Kennes2018strong,Isobe2018, You2018} as well as phonon mechanism \cite{Wu2018phonon}. Theory presented in this paper builds upon the theoretical framework developed in our previous work \cite{Wu2018phonon}, where phonons mediate pairing, while our qualitative results for $d$-wave channel should be largely independent of the exact pairing mechanisms. The continuum model that we employ captures the sublattice and layer dependence of the moir\'e electronic wave function, which is crucial for our findings. 

\begin{figure*}[t]
	\includegraphics[width=2\columnwidth]{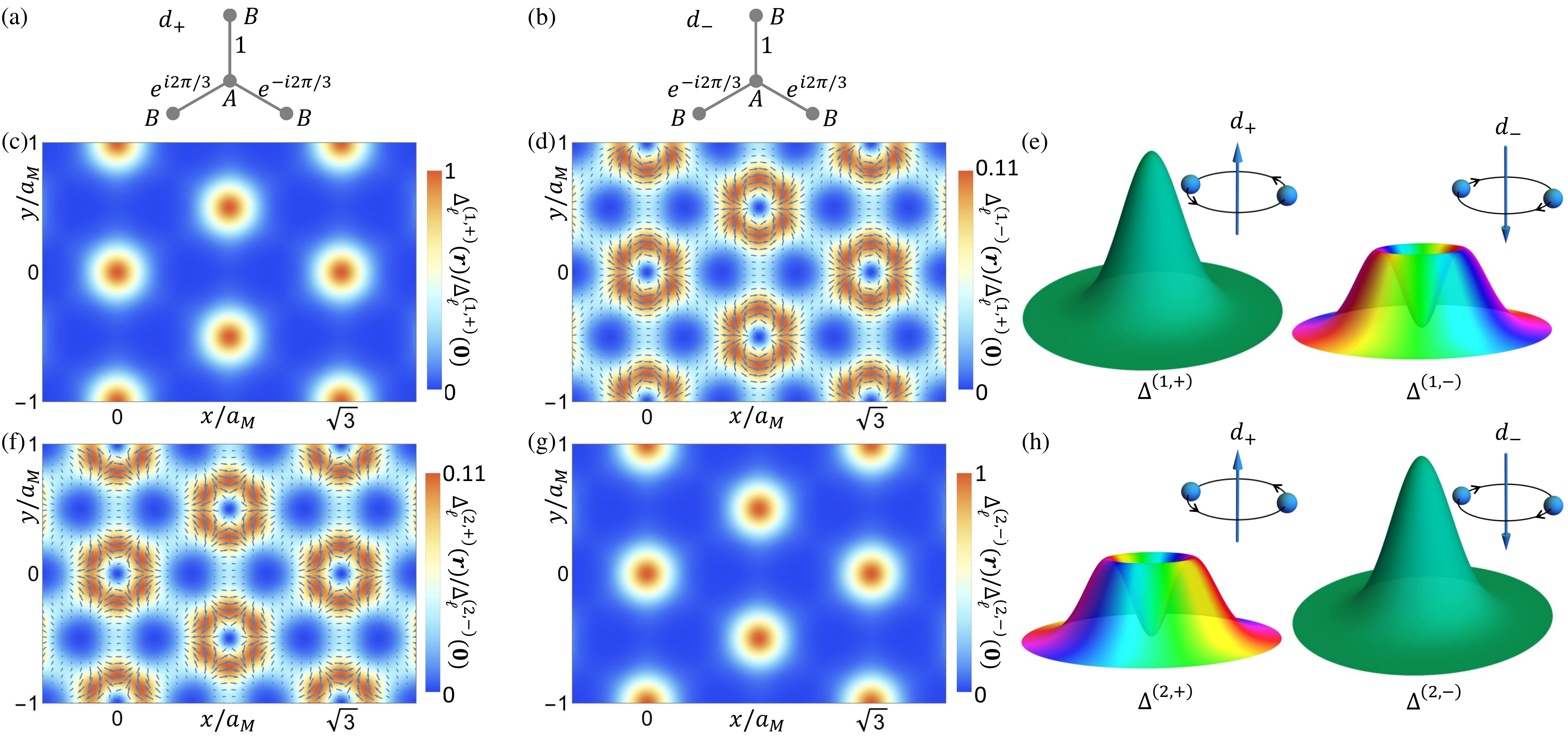}
	\caption{(a) $d_{+}$ and (b) $d_{-}$  pairings at the atomic scale, where electrons 
		at nearest-neighbor sites are paired with the indicated bond-dependent phase factors. (c) and (d) respectively show the pair amplitudes $\Delta_{\ell}^{(1, +)}(\rr)$ and $\Delta_{\ell}^{(1, -)}(\rr)$ in $\hat{\Gamma}_1$. In (c), $\Delta_{\ell}^{(1, +)}(\rr)$ normalized by its value at $\rr=0$ (AA region center) is real. In (d), $\Delta_{\ell}^{(1, -)}(\rr)$ also normalized by $\Delta_{\ell}^{(1, +)}(\boldsymbol{0})$ is complex, and its magnitude and phase are indicated respectively by the color scale and the vectors. (e) Schematic plots of the $\hat{\Gamma}_1$ pair amplitudes around $\rr=0$.  (f), (g) and (h) are corresponding plots for pair amplitudes in $\hat{\Gamma}_2$.
}
	\label{Fig:pair_potential}
\end{figure*}

We demonstrate two  phenomena for chiral $d$-wave states in moir\'e superlattice, i.e., spontaneous vortices in the pairing order parameters and spontaneous bulk supercurrent. The presence of spontaneous vortices can be anticipated by examining the angular momentum of a Cooper pair, which is explained briefly in the following and thoroughly in Sec.~\ref{sec:dpairing}. In the $d$-wave pairing, the relative motion (RM) of two electrons in a Cooper pair has either $d+id$ ($d_+$) or $d-id$ ($d_-$) symmetry, which carry opposite angular momenta under the transformation of a three-fold rotation. The total angular momentum of a Cooper pair has contributions from both the RM and the center-of-mass motion (COMM). Due to the enlarged moir\'e superlattice, COMM can also carry a finite angular momentum  while preserving the moir\'e periodicity. By matching the total angular momentum, Cooper pairs with $d_+$ and $d_-$ RMs are intrinsically coupled. Overall, there are still two independent chiral pairing channels, which are respectively  labeled as $\hat{\Gamma}_1$ and $\hat{\Gamma}_2$ [see Eq.~(\ref{Gamma12}) for definition]. In $\hat{\Gamma}_1$, COMM associated with $d_+$ RM has an $s$-wave symmetry, while COMM for $d_-$ RM has a spontaneous vortex-antivortex lattice configuration. The other channel $\hat{\Gamma}_2$ is the time-reversal counterpart of $\hat{\Gamma}_1$. The above order parameter structures are  illustrated in Fig.~\ref{Fig:pair_potential}. We note that  Ref.~\onlinecite{Lin2018} also reported spontaneous vortices in the superconductivity order parameters, but for a mixed $d$ and $p$ wave pairing state. In this paper we explain the origin of vortices based on the angular momentum of Cooper pairs.

The chiral $d$-wave {\it ground} state carries spontaneous circulating supercurrent in the bulk. This is possible because each moir\'e unit cell contains a large number of atomic sites that support current flow. We find that supercurrent has one component circulating around $\hat{z}$ axis (perpendicular to TBLG) and another component circulating between the two graphene layers in TBLG. The supercurrent distribution pattern is characterized by both a magnetic dipole moment and a magnetic toroidal dipole moment.

We also discuss the gap structure in the superconducting states.  The chiral $d$-wave states are gapped and also topological as characterized by an integer Chern number. The two-component $d$-wave pairing channels can also lead to nematic states, which break rotation symmetry but preserve time-reversal symmetry.  Nematic $d$-wave superconductors are gapless with point nodes. Within weak-coupling mean-field theory, chiral $d$-wave states are energetically more favored. However, nematic $d$-wave states could be stabilized near the critical temperature when the six-fold rotational symmetry of the TBLG is broken, for example by uniaxial strain.

\begin{figure}[t]
	\includegraphics[width=1\columnwidth]{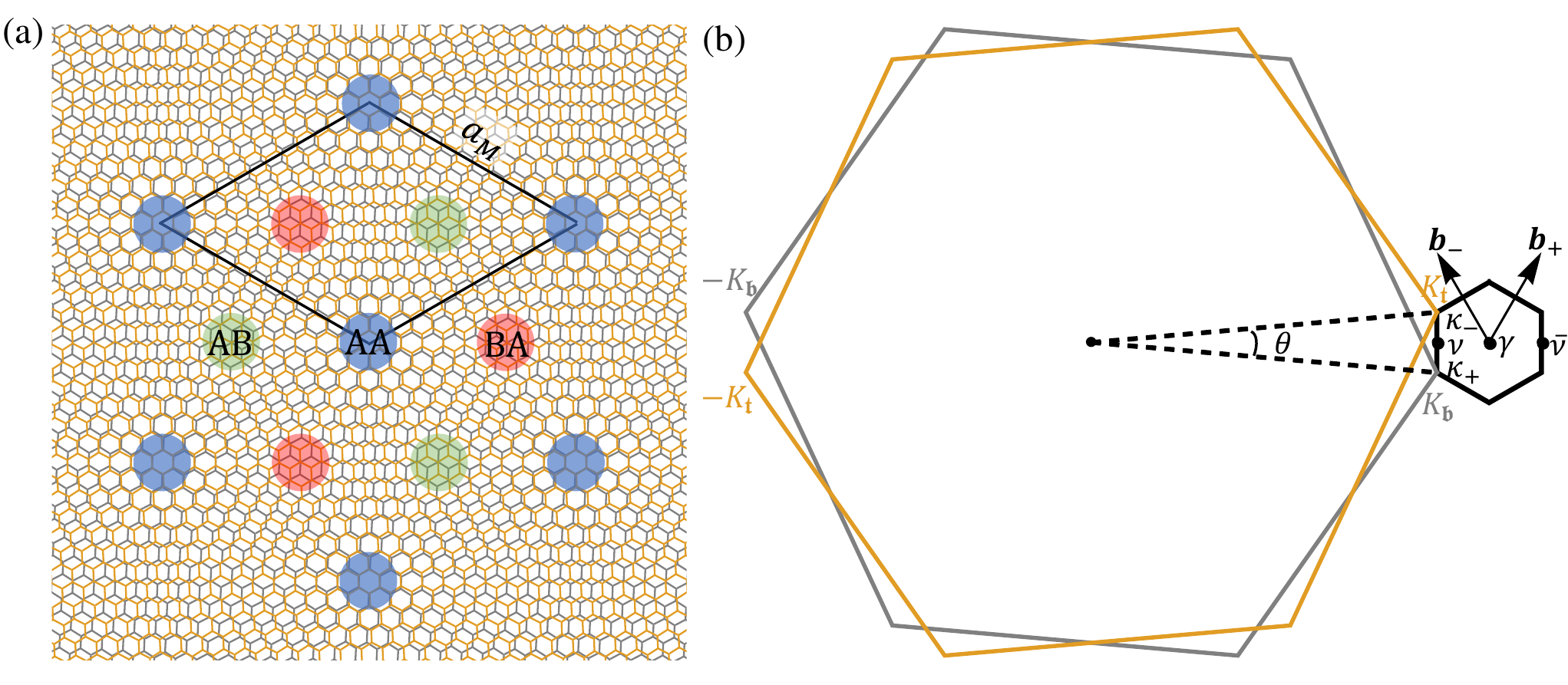}
	\caption{(a) Moir\'e pattern in TBLG. (b) The small black hexagon represents the moir\'e Brillouin zone, while the gray and yellow hexagons show the Brillouin zone associated with bottom and top layers.}
	\label{Fig:moire_pattern}
\end{figure}

This paper is organized as follows. Section~\ref{sec:band} sets up the single-particle moir\'e Hamiltonian. In Sec.~\ref{sec:dpairing}, we study $d$-wave pairing within mean-field theory, present the critical temperature, and discuss the superconductivity order parameters, including the spontaneous vortices. In Sec.~\ref{sec:supercurrent}, we illustrate the spontaneous supercurrent in the chiral $d$-wave state. Sections ~\ref{Sec:TopChiral} and ~\ref{Sec:NematicD} respectively present the gap structure of chiral and nematic $d$-wave states. In Sec.~\ref{Sec:TopChiral}, we also show that the chiral $d$-wave state is topological by computing the Berry curvature and the Chern number. Finally, a brief discussion and summary are given in Sec.~\ref{sec:discussion}.

\section{Moir\'e bands}
\label{sec:band}

\begin{figure}[t]
	\includegraphics[width=1\columnwidth]{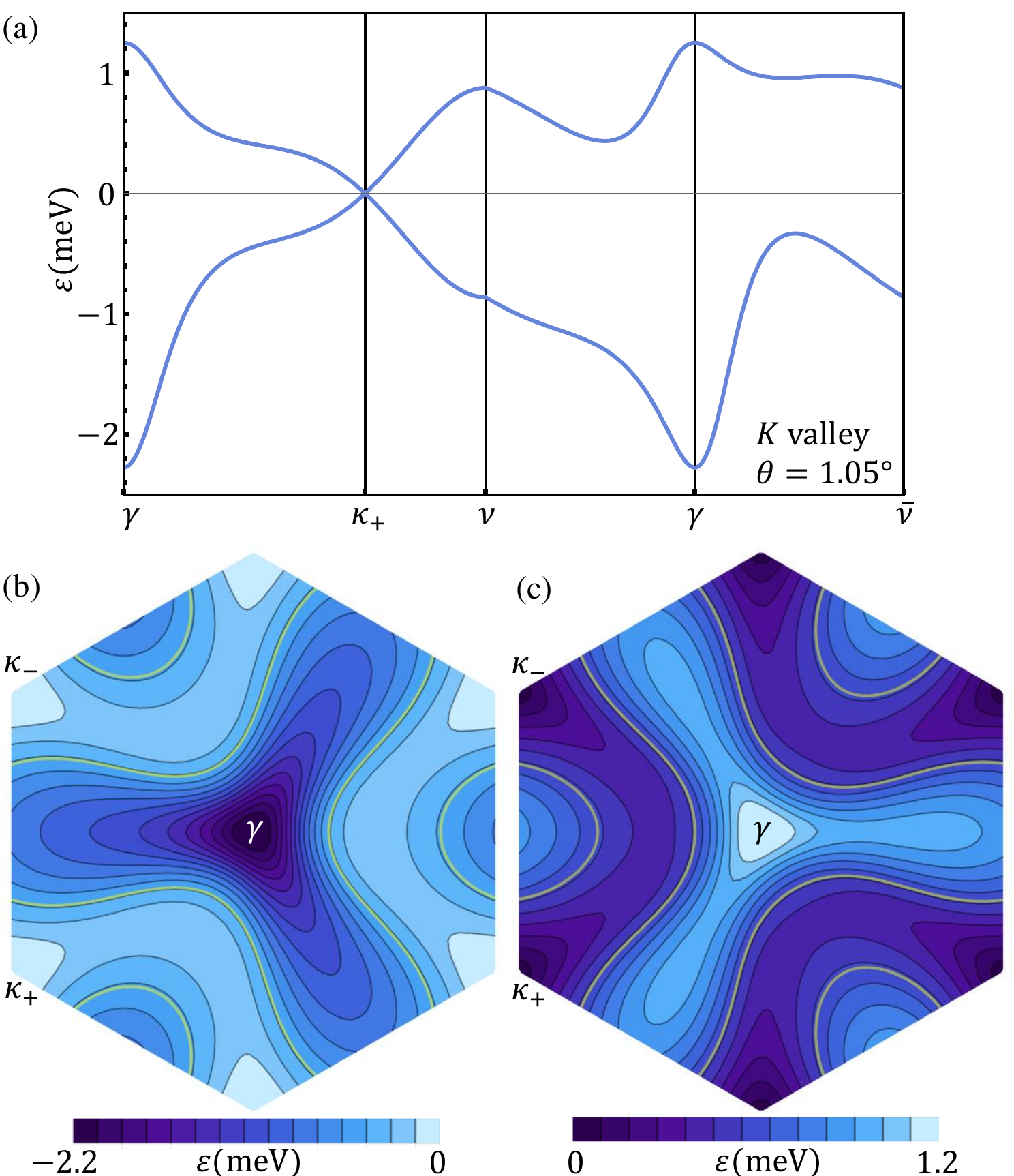}
	\caption{(a) Moir\'e band structure along high symmetry lines for $+K$ valley and $\theta=1.05^{\circ}$. Only the two bands close to zero energy (set by the Dirac point energy) are shown. Energy contour plots for the lower and upper bands in (a) are shown respectively in (b) and (c). The yellow contours in (b) and (c) indicate the Fermi surface when the lower or upper band is half filled.}
	\label{Fig:moire_bands}
\end{figure}

We construct TBLG with point group $D_6$ by starting from AA stacked bilayer graphene, and then
rotate the bottom and top layers by angles $-\theta/2$ and $+\theta/2$ around one of the hexagon center, as illustrated in Fig.~\ref{Fig:moire_pattern}(a). The origin of coordinates is chosen to be on this
rotation axis and half-way between layers.
The $D_6$ point group symmetry with respect to this origin   is generated by a sixfold rotation $\hat{C}_6$
around the $\hat{z}$ axis, and twofold rotations  $\hat{\mathcal{M}}_x$ and
$\hat{\mathcal{M}}_y$ respectively around the $\hat{x}$ and $\hat{y}$ axes.
The operations $\hat{\mathcal{M}}_{x, y}$ swap the two layers.
Because spin-orbit interactions
are negligible in graphene \cite{Brataas2006,Min2006},
electrons have accurate spin SU(2) symmetry. Therefore, superconductivity can be classified as spin singlet and triplet.

At a small twist angle $\theta$, TBLG has a triangular moir\'e pattern with a long period $a_M=a_0/[2\sin(\theta/2)]$, where $a_0$ is the lattice constant of monolayer graphene. In the moir\'e pattern, there are three notable regions, where the local interlayer coordinations are of AA, AB and BA types, as highlighted in Fig.~\ref{Fig:moire_pattern}(a).

The single-particle physics of TBLG with small $\theta$ can be
described using a continuum moir\'e Hamiltonian, in which the atomic-scale commensurability plays no role. The moir\'e Hamiltonian \cite{Bistritzer2011} is spin-independent and
is given in valley $\tau K$ by
\begin{equation}
\mathcal{H}_{\tau}=\begin{pmatrix}
h_{\tau \mathfrak{b}}(\kk) & T_{\tau }(\rr) \\
T^{\dagger}_{\tau}(\rr) & h_{\tau \mathfrak{t}}(\kk)
\end{pmatrix},
\label{Hmoire}
\end{equation}
where $\tau=\pm$ is the valley index.
$h_{\tau \mathfrak{b}}$ and $h_{\tau \mathfrak{t}}$ are the Dirac Hamiltonians of the bottom ($\mathfrak{b}$) and top ($\mathfrak{t}$) layers
\begin{equation}
h_{\tau \ell}(\kk) = e^{-i\tau \ell \frac{\theta}{4} \sigma_z }[\hbar v_F (\kk-\tau \boldsymbol{\kappa}_{\ell})\cdot (\tau \sigma_x, \sigma_y)]e^{+i\tau \ell \frac{\theta}{4} \sigma_z},
\end{equation}
where $\ell$ is $+1$ ($-1$) for the $\mathfrak{b}$ ($\mathfrak{t}$) layer,
$v_F$ is the bare Dirac velocity($\sim 10^6$ m/s), and
$\sigma_{x, y}$ are Pauli matrices that act in the sublattice space.
Because of the rotation, the Dirac cone position in layer $\ell$ and valley $\tau$ is shifted to
$\tau \boldsymbol{\kappa}_{\ell}$.
We choose a moir\'e Brillouin zone (MBZ) in which  $\boldsymbol{\kappa}_{\ell}$ is
located at the  corners, and refer to the
MBZ center below as the $\gamma$ point.
$\boldsymbol{\kappa}_{\ell}$ is then given by $[4\pi/(3 a_M)](-\sqrt{3}/2, -\ell/2)$.
The interlayer tunneling terms are sublattice-dependent and vary periodically with  the real space position $\rr$
\begin{equation}
T_{\tau}(\rr)=
T_{\tau}^{(0)}
+e^{-i \tau \bb_+ \cdot \rr} T_{\tau}^{(+1)}
+e^{-i \tau \bb_- \cdot \rr} T_{\tau}^{(-1)}
\end{equation}
where $\bb_{\pm}$ are moir\'e reciprocal lattice vectors given by $[4\pi/(\sqrt{3} a_M)](\pm1/2, \sqrt{3}/2)$ and
$T_{\tau}^{(j)} = w_0 \sigma_0 + w_1 \cos(2\pi j/3)\sigma_x+ \tau w_1\sin(2\pi j/3) \sigma_y$. Here $w_0$ and $w_1$ are parameters that respectively determine the tunneling in AA and AB/BA regions. $w_0$ and $w_1$ are different because the interlayer distance in the AA region is larger 
than that in AB/BA regions, and therefore, $|w_0|<|w_1|$ is expected. We take $w_0 = 90$ meV and $w_1 = 117 $ meV from Ref.~\onlinecite{Jung2014}. 
For this choice of parameters, the largest magic angle, at which the Dirac velocity reaches a minimum value, is about $1.025^{\circ}$. Near this magic angle, the two nearly flat bands close to zero energy (set by the Dirac point energy) are separated from higher or lower energy bands by a gap of about 35 meV, which is consistent with experiments \cite{Cao2018Magnetic,Cao2018Super} and motivates the use of the two tunneling parameters. Fig.~\ref{Fig:moire_bands} shows the moir\'e band structure at $\theta=1.05^{\circ}$, which will be used below as a representative example for the discussion of superconductivity properties.

We note that the moir\'e Hamiltonian builds in the $D_6$ point group symmetry and also the time-reversal symmetry $\hat{\mathcal{T}}$ , as $h_{\tau \ell}^*(\kk)=h_{(-\tau) \ell}(-\kk)$ and $T^*_{\tau}(\rr)=T_{-\tau}(\rr)$. The $\hat{\mathcal{T}}$  symmetry implies that $\varepsilon_{\tau}(\qq)=\varepsilon_{-\tau}(-\qq)$, where $\varepsilon_{\tau}$ is the band energy in valley $\tau K$ and $\qq$ is the momentum relative to the $\gamma$ point. The band structure within one valley has strong trigonal warping as demonstrated in Fig.~\ref{Fig:moire_bands}, and therefore, $\varepsilon_{\tau}(\qq) \neq \varepsilon_{\tau}(-\qq)$. Because of this feature in the band structure, intervalley electron pairing is more favored.

\section{$d$-wave pairings}
\label{sec:dpairing}
We  studied the coupling between moir\'e electrons and in-plane optical phonon modes associated with each 
graphene layer in Ref.~\onlinecite{Wu2018phonon}, and found that this coupling mediates the following intervalley electron pairing interactions
\begin{equation}
\begin{aligned}
H_{\text{BCS}}=-4 \int d^2 \rr \{ g_{E_2}[\hat{\psi}^{\dagger}_{+ A \ell s}\hat{\psi}^{\dagger}_{- A \ell s'}\hat{\psi}_{- B \ell s'}\hat{\psi}_{+ B \ell s}+h.c.]\\+g_{A_1}[\hat{\psi}^{\dagger}_{+ A \ell  s}\hat{\psi}^{\dagger}_{- A \ell s'}\hat{\psi}_{+ B \ell s'}\hat{\psi}_{- B \ell s}+ h.c. ]
\\+g_{A_1}[\hat{\psi}^{\dagger}_{+ A \ell s}\hat{\psi}^{\dagger}_{- B \ell s'}\hat{\psi}_{+ A \ell s'}\hat{\psi}_{- B \ell s}+ (A \leftrightarrow B) ]
\},
\end{aligned}
\label{HBCS}
\end{equation}
where the operators $\hat{\psi}^{\dagger}$ and $\hat{\psi}$ are at the same coarse-grained position $\rr$.
The subscript $\pm$ distinguish the two valleys, $A$ and $B$ label sublattices, $\ell$ refers to layers, and $s$ and $s'$ are spin indices. In Eq.~(\ref{HBCS}), interactions in the first line are induced by phonon modes near $\Gamma$ point, and those in the second and third lines are generated by phonon modes near $\pm K$ points.  The coupling constants $g_{E_2}$ and $g_{A_1}$ are respectively about  52 and 69 meV$\cdot$nm$^2$.

In $H_{\text{BCS}}$, there are two distinct spin-singlet pairing channels: (i) intrasublattice pairing, e.g., $\epsilon_{s s' }\hat{\psi}^{\dagger}_{+ A \ell s}\hat{\psi}^{\dagger}_{- A \ell s'}$ and (ii) intersublattice pairing, e.g., $\epsilon_{s s' } \hat{\psi}^{\dagger}_{+ A \ell s}\hat{\psi}^{\dagger}_{- B \ell s'}$, where $\epsilon$ is the fully antisymmetric tensor with $\epsilon_{\uparrow \downarrow}=1$. While the intrasublattice pairing channels are generated by both $\Gamma$ and $\pm K$ phonons,
only the  $\pm K$ phonons contribute to intersublattice pairing. The intra- and intersublattice pairings respectively  have $s$-wave and $d$-wave symmetries because
electrons at different sublattices and opposite valleys share the {\it same} angular momentum under the threefold rotation
$\hat{C}_3 \hat{\psi}^{\dagger}(\rr) \hat{C}_3^{-1} = \exp[i 2\pi \sigma_z \tau_z /3] \hat{\psi}^{\dagger}(\mathcal{R}_3 \rr) $, where $\mathcal{R}_3$ is the real-space three-fold rotational matrix.   Intersublattice pairings $\hat{P}_{\ell +}(\rr)=\epsilon_{s s' } \hat{\psi}^{\dagger}_{+ A \ell s}(\rr)\hat{\psi}^{\dagger}_{- B \ell s'}(\rr)$ and $\hat{P}_{\ell -}(\rr)=\epsilon_{s s' } \hat{\psi}^{\dagger}_{+ B \ell s}(\rr)\hat{\psi}^{\dagger}_{- A \ell s'}(\rr)$ carry opposite angular momenta ($\pm 2$),
\begin{equation}
\begin{aligned}
\hat{C}_3 \hat{P}_{\ell +}(\rr) \hat{C}_3^{-1}
&= e^{+i \frac{4\pi}{3}} \hat{P}_{\ell +}(\mathcal{R}_3\rr),\\
\hat{C}_3 \hat{P}_{\ell -}(\rr) \hat{C}_3^{-1}
&= e^{-i \frac{4\pi}{3}} \hat{P}_{\ell -}(\mathcal{R}_3\rr).
\end{aligned}
\label{RMs}
\end{equation}
Here the angular momentum is defined based on the $\hat{C}_3$ operation and therefore is determined up to modulo 3.    $\hat{P}_{\ell +}$ and $\hat{P}_{\ell -}$  correspond to
chiral $d_+$ and $d_-$ pairings, respectively.   At the atomic scale, chiral $d$-wave pairings are realized by forming
nearest-neighbor spin-singlet Cooper pairs with bond-dependent phase factors, as illustrated in
Figs.~\ref{Fig:pair_potential}(a, b).  
The opposite angular momenta associated with $\hat{P}_{\ell \pm}$  arise from the relative motion between two electrons in one Cooper pair.

We focus on the $d$-wave pairing, assuming that the $s$-wave pairing is suppressed by Coulomb repulsion effects.
In Ref.~\onlinecite{Wu2018phonon}, the  $d_+$ and $d_-$ pairings are considered to be independent, which is a good approximation when estimating the critical temperature. Here we present the full theory and show that $d_+$ and $d_-$ pairings are coupled in the linearized gap equation, although only weakly.

We perform mean-field theory, and the local pair amplitude is given by
\begin{equation}
\begin{aligned}
\Delta_{\ell}^{(+)}(\rr) &= \langle \hat{\psi}_{- B \ell \downarrow}(\rr)  \hat{\psi}_{+ A \ell \uparrow}(\rr) \rangle = - \langle \hat{\psi}_{- B \ell \uparrow}(\rr)  \hat{\psi}_{+ A \ell \downarrow}(\rr) \rangle,\\
\Delta_{\ell}^{(-)}(\rr) &= \langle \hat{\psi}_{- A \ell \downarrow}(\rr)  \hat{\psi}_{+ B \ell \uparrow}(\rr) \rangle = - \langle \hat{\psi}_{- A \ell \uparrow}(\rr)  \hat{\psi}_{+ B \ell \downarrow}(\rr) \rangle.
\end{aligned}
\end{equation}
We further assume that the pair amplitude has
moir\'e periodicity and can be expressed using harmonic expansion
$\Delta_{\ell}^{(d)}(\rr) = \sum_{\bb} e^{i \bb \cdot \rr} \Delta_{\bb,\ell}^{(d)}$, where the superscript $d=\pm$ represents the two $d$-wave pairings and $\bb$ is the moir\'e reciprocal lattice vectors. The linearized gap equation is given by
\begin{equation}
\begin{aligned}
\Delta^{(d)}_{\bb,\ell} = & \sum_{\bb'\ell' d'} \chi^{(\bb \ell d) }_{(\bb' \ell' d')} \Delta^{(d')}_{\bb',\ell'}, \\
\chi^{(\bb \ell d) }_{(\bb' \ell' d')}= & \frac{ 4 g_{A_1}}{\mathcal{A}} \sum_{\qq,n_1,n_2}  \Big\{  \frac{1-n_F[\varepsilon_{n_1}(\qq)]-n_F[\varepsilon_{n_2}(\qq)]}{\varepsilon_{n_1}(\qq)+\varepsilon_{n_2}(\qq)-2\mu} \\
& \times [\langle u_{n_1}(\qq) | \sigma_{d} | u_{n_2}(\qq) \rangle_{\bb, \ell}]^* \\
& \times\langle u_{n_1}(\qq) | \sigma_{d'} | u_{n_2}(\qq) \rangle_{\bb', \ell'} \Big\},
\end{aligned}
\label{chi}
\end{equation}
where $\mathcal{A}$ is the system area, $\qq$ is a momentum within moir\'e Brillouin zone, $n_{1, 2}$ are moir\'e band labels in $+K$ valley for one spin component, $\varepsilon_n$ and $|u_n\rangle$ are the corresponding energies and wave functions,
$n_F(\varepsilon)$ is the Fermi-Dirac occupation function, and $\mu$ is the chemical potential. The band energy  $\varepsilon_n$ is measured relative to the Dirac point. The overlap function $\langle ... \rangle_{\bb, \ell}$ is the layer-resolved matrix element of the combined operator $\exp(i \bb \cdot \rr )\sigma_{\pm}$, where $\sigma_{\pm}=(\sigma_x \pm i \sigma_y)/2$. Note that the time-reversal symmetry of the moir\'e Hamiltonian has been employed to simplify (\ref{chi}).

The operator $\sigma_{\pm}$ is closely related to the velocity operator
$\hbar  \hat{\bv}_{\tau}=\partial \mathcal{H}_{\tau}/\partial \kk $.
Near the magic angle, the velocity of the flat bands is strongly suppressed, but
the layer counterflow velocity, which is approximately determined by the operator $ \ell \sigma_{\pm} $,  remains large \cite{Bistritzer2011}.
As a result, the leading $d$-wave instability has pair amplitudes of opposite signs in the two layers
$\Delta_{\mathfrak{b}}^{(d)}(\rr) = - \Delta_{\mathfrak{t}}^{(d)}(\rr)$.

\begin{figure}[t!]
	\includegraphics[width=0.8\columnwidth]{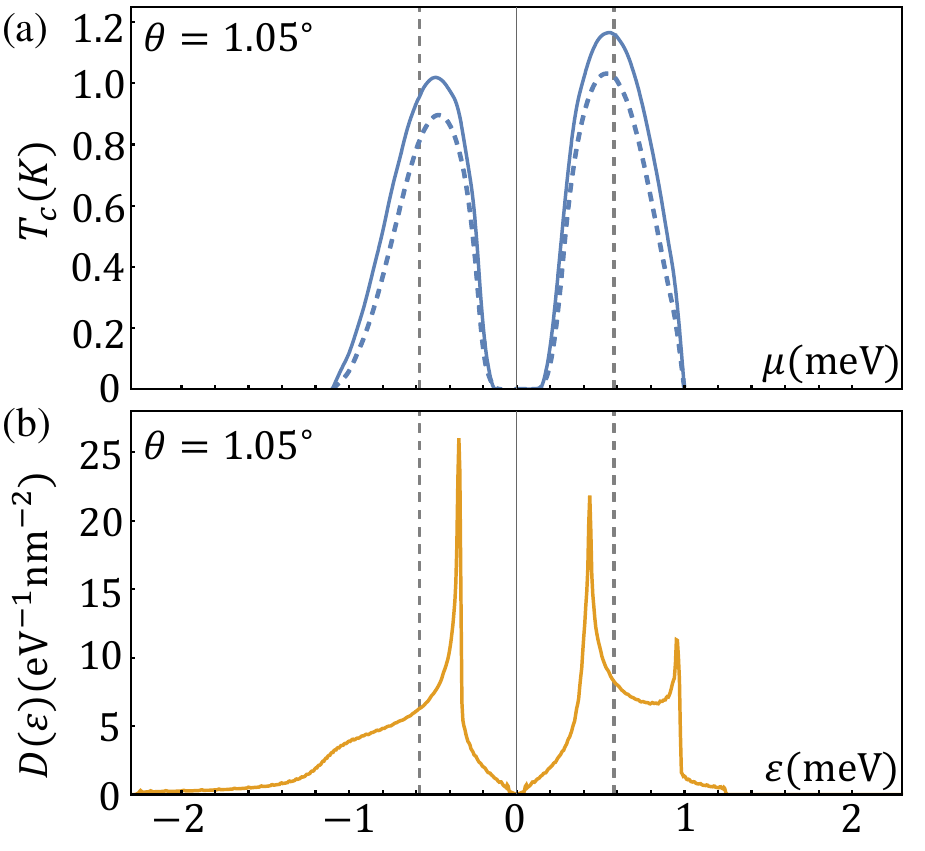}
	\caption{(a) Critical temperature $T_c$ for $d$-wave pairing as a function of the chemical potential $\mu$. The solid and dashed lines are obtained with and without considering the coupling between $d_+$ and $d_-$ pairings. (b) Density of states per spin and per valley as a function of energy. The vertical dashed lines in (a) and (b) show the chemical potential at which the lower or upper flat band is half filled. The twist angle is $1.05^{\circ}$.}
	\label{Fig:Tc_d}
\end{figure}

The superconductivity critical temperature $T_c$ is obtained by requiring that the largest eigenvalue of the pair susceptibility $\chi$ is equal to 1.  In Fig.~\ref{Fig:Tc_d}(a), we show the theoretical $T_c$ as a function of chemical potential $\mu$ for twist angle $\theta=1.05^{\circ}$. The trend of $T_c(\mu)$ does not exactly follow the density of states shown in Fig.~\ref{Fig:Tc_d}(b), because all states in the nearly flat band can effectively contribute to the pairing. $T_c$ vanishes at the Dirac point energy ($\mu=0$), and peaks near the chemical potential at which the lower or upper flat band is half filled. These features of $T_c(\mu)$ are in qualitative agreement with experiments.\cite{Cao2018Super, Dean2018Tuning} The maximum $T_c$ in Fig.~\ref{Fig:Tc_d}(a) is about 1.2 K, comparable to the experimental values. We note that theoretical $T_c$ depends on the flatness of the moir\'e bands, and therefore, on the model parameters, which are not known precisely.  In the calculation, $\chi$ is computed by including momenta $\bb$ up to the third moir\'e reciprocal lattice vector shell and by retaining only the two flat bands near zero energy because of their high density of states.

The largest eigenvalue of $\chi$ corresponds to two degenerate eigenvectors $\lambda_1$ and $\lambda_2$, where the degeneracy is protected by the point group symmetries and also time-reversal symmetry.  
The pair amplitudes $[\Delta_{\ell}^{(j, +)}(\rr), \Delta_{\ell}^{(j,-)}(\rr)]$ associated with $\lambda_j$ ($j=1$ and 2) lead to the following mean-field pair potential
\begin{equation}
\begin{aligned}
\hat{\Gamma}_j&=-4 g_{A_1}\int d\rr \hat{\Gamma}_j(\rr),\\
\hat{\Gamma}_j(\rr)&=\sum_{\ell}  
\Delta_{\ell}^{(j, +)}(\rr) \hat{P}_{\ell +}(\rr) +  
\Delta_{\ell}^{(j, -)}(\rr) \hat{P}_{\ell -}(\rr).
\end{aligned}
\label{Gamma12}
\end{equation}
We distinguish $\hat{\Gamma}_1$ and $\hat{\Gamma}_2$  by the $\hat{C}_3$ rotational symmetry
\begin{equation}
\hat{C}_3 \hat{\Gamma}_1 \hat{C}_3^{-1}= e^{i 4\pi/3}\hat{\Gamma}_1,\,\,\,\,\,\,\,\,\,
\hat{C}_3 \hat{\Gamma}_2 \hat{C}_3^{-1}= e^{-i 4\pi/3}\hat{\Gamma}_2,
\label{PairPotential}
\end{equation}
which is realized by requiring that
\begin{equation}
\begin{aligned}
&\Delta_{\ell}^{(1, +)}(\mathcal{R}_3\rr)=\Delta_{\ell}^{(1, +)}(\rr), \\
&\Delta_{\ell}^{(1, -)}(\mathcal{R}_3\rr)=e^{-i\frac{2\pi}{3}}\Delta_{\ell}^{(1, -)}(\rr), \\
&\Delta_{\ell}^{(2, +)}(\mathcal{R}_3\rr)=e^{+i\frac{2\pi}{3}}\Delta_{\ell}^{(2, +)}(\rr),\\ 
&\Delta_{\ell}^{(2, -)}(\mathcal{R}_3\rr)=\Delta_{\ell}^{(2, -)}(\rr).
\end{aligned}
\label{COMM}
\end{equation}
We interpret the pair amplitude $\Delta_{\ell}^{(j, \pm)}$ as the envelope wave function for the COMM of the Cooper pair. In the linearized gap equation, Cooper pairs with relative motions $d_+$ and $d_-$ are coupled by adjusting their  COMMs according to Eq.~(\ref{COMM}).  The COMM can carry a finite angular momentum without breaking the moir\'e periodicity. The total angular momentum of the pair potential $\hat{\Gamma}_j$ is contributed by both the RM and the COMM. 
The pair amplitudes $[\Delta_{\ell}^{(1, +)}(\rr), \Delta_{\ell}^{(1,-)}(\rr)]$ in $\hat{\Gamma}_1$ are shown in Figs.~\ref{Fig:pair_potential}(c, d, e). $\Delta_{\ell}^{(1, +)}(\rr)$ has an $s$-wave symmetry and peaks near AA regions following the electron density distribution, while $\Delta_{\ell}^{(1, -)}(\rr)$ has a vortex-antivortex lattice configuration, in which vortices centered around AA, AB and BA regions have vorticity $+2$, $-1$ and $-1$ respectively. Therefore, $\Delta_{\ell}^{(1, -)}(\rr)$ satisfies (\ref{COMM}) and preserves the moir\'e periodicity because of the zero total vorticity. As shown in Figs.~\ref{Fig:pair_potential}(c, d, e),  $\Delta_{\ell}^{(1, +)}$ is dominant in $\hat{\Gamma}_1$, and the vortices in $\Delta_{\ell}^{(1, -)}$  can be regarded as a secondary effect. Figs.~\ref{Fig:pair_potential}(f, g, h) show the pair amplitudes in $\hat{\Gamma}_2$, which is the time reversal partner of $\hat{\Gamma}_1$.

We also calculate $T_c$ by neglecting the coupling between $d_+$ and $d_-$ relative motions, and the maximum $T_c$ calculated in this way is slightly lower than that obtained from the full calculation [Fig.~\ref{Fig:Tc_d}(a)], which is consistent with the fact that the vortices are only a perturbative effect.

As a side remark, we note that the periodic modulation of the pair amplitude is actually a ubiquitous phenomenon in crystalline superconductors.
We can take superconducting aluminum as an example, which has a long coherence length ($\sim 1600 $ nm). However, the $s$-wave pair amplitude $\Delta(\rr)=\langle \psi_\downarrow(\rr) \psi_\uparrow(\rr)\rangle$ in aluminum  has the lattice periodicity and varies within one unit cell following the variation of the normal state electron wave function \cite{walter1973electronic}. This variation of the pair amplitude in the superconducting  {\it ground} state is typically a negligible effect, because it is a modulation over a very short distance determined by the lattice constant (0.4 nm in the case of aluminum). This modulation becomes noticeable in moir\'e pattern because of the large moir\'e period ($\sim$ 13.4 nm for $\theta =1.05^{\circ}$). Cuprates present another context in which the spatial modulation of the pair amplitude is important. In the $d_{x^2-y^2}$ pairing state of cuprates, the pair amplitudes along the $\hat{x}$ and $\hat{y}$ bonds within one unit cell are phase shifted by $\pi$.

The pair potentials $\hat{\Gamma}_1$ and $\hat{\Gamma}_2$ form a two-dimensional $E_2$ representation of the $D_6$ point group, and lead to chiral $d$-wave superconductivities, which are time-reversal symmetry breaking and fully gapped as discussed in Sec.~\ref{Sec:TopChiral}. Linear combinations of $\hat{\Gamma}_1$ and $\hat{\Gamma}_2$ can give rise to nematic $d$-wave superconductivities, which are time-reversal symmetric but break rotational symmetries. Nematic states are gapless as discussed in Sec.~\ref{Sec:NematicD}. Therefore, chiral $d$-wave superconductivities should be favored over nematic states within  mean-field theory considered in this paper.

\begin{figure*}[t]
	\includegraphics[width=1.6\columnwidth]{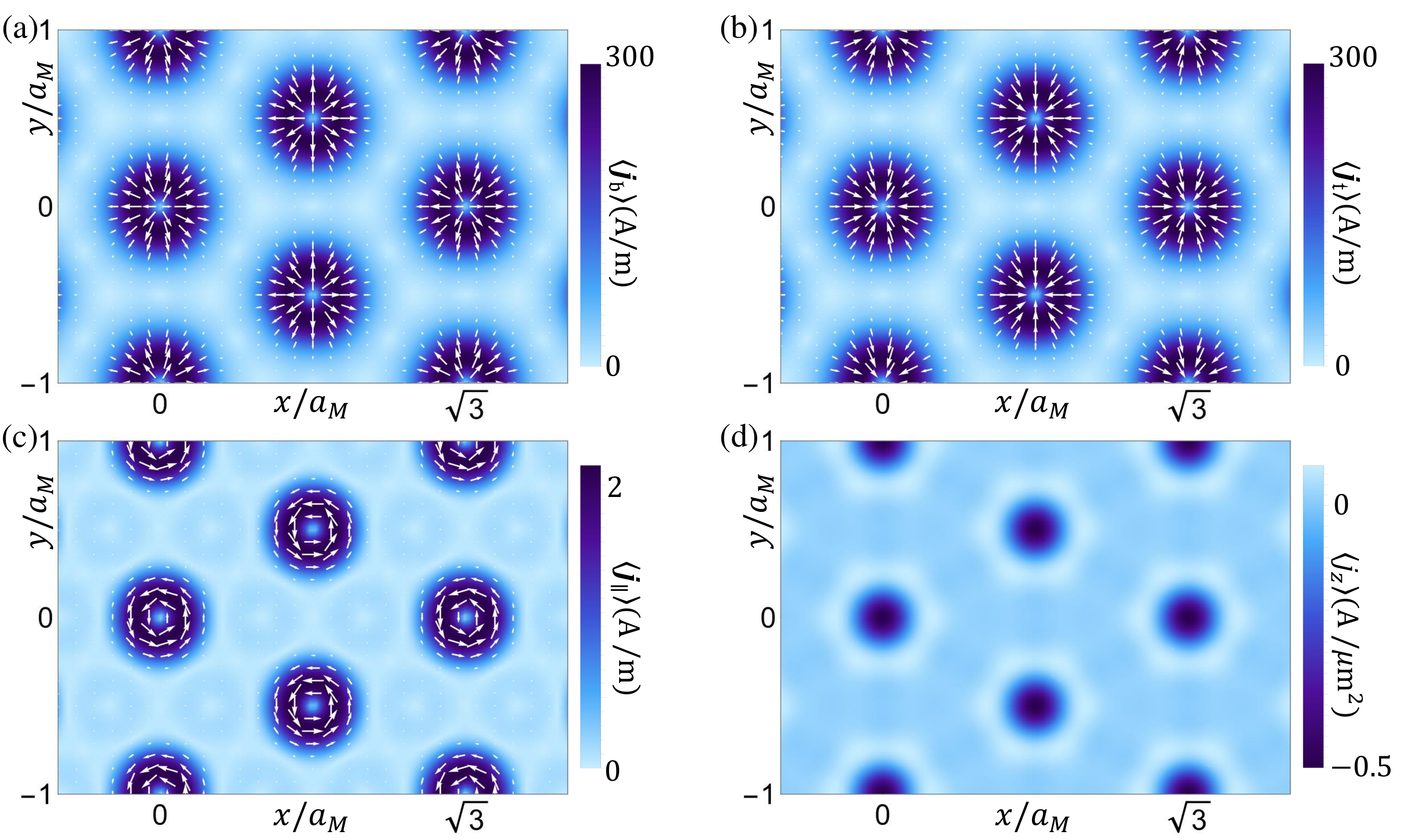}
	\caption{In-plane current flow pattern for (a) bottom layer, (b) top layer and (c) the total. The color scale shows the magnitude, while the vectors indicate the flow direction. (d) Interlayer tunneling current density. The calculation is based on the perturbation theory for the chiral $d$-wave state with $\hat{\Gamma}_1$ pair potential. The chemical potential used in the calculation corresponds to half-filling of the lower-flat band. The pair amplitudes are shown in Figs.~\ref{Fig:pair_potential}(c,d) with the overall scale $-4 g_{A_1} \Delta_{\mathfrak{b}}^{(1,+)}(\boldsymbol{0})$ set to be 1 meV. The  magnitude of the current is proportional to $|\Delta_{\mathfrak{b}}^{(1,+)}(\boldsymbol{0})|^2$ in the perturbation theory.  }
	\label{Fig:supercurrent}
\end{figure*}

\section{Spontaneous supercurrent}
\label{sec:supercurrent}
We demonstrate that the chiral $d$-wave states support spontaneous bulk supercurrent. We first present various current operators. The in-plane current density operator for layer $\ell$ is given by
\begin{equation}
\begin{aligned}
\jj_{\ell}(\rr)=(-e_0)\sum_{\tau, s} \hat{\psi}_{\tau \ell s}^{\dagger}(\rr) \boldsymbol{\mathcal{J}}_{\tau \ell} \hat{\psi}_{\tau \ell s}(\rr),
\end{aligned}
\end{equation}
where $-e_0$ is the electron charge ($e_0>0$) and  $\boldsymbol{\mathcal{J}}_{\tau \ell}$ is the velocity operator that acts in the sublattice space
\begin{equation}
\begin{aligned}
\boldsymbol{\mathcal{J}}_{\tau \ell}=\frac{\partial h_{\tau\ell}(\kk)}{\hbar \partial \kk} 
=v_F e^{-i\tau \ell \frac{\theta}{4} \sigma_z }(\tau \sigma_x, \sigma_y)e^{+i\tau \ell \frac{\theta}{4} \sigma_z}.
\end{aligned}
\end{equation} 
Note that $\boldsymbol{\mathcal{J}}_{\tau \ell}$ derived from the Dirac Hamiltonian is independent of momentum and position. $\jj_{\ell}(\rr)$ is a two-component vector representing the in-plane current flow in layer $\ell$, and has the unit of current per length. The total in-plane current $\jj_{\parallel}$ is the sum of $\jj_{\mathfrak{b}}$ and $\jj_{\mathfrak{t}}$.

The interlayer tunneling leads to the following out-of-plane current density operator
\begin{equation}
j_{z}(\rr)=(-e_0)\frac{i}{\hbar}\sum_{\tau,s}[\hat{\psi}_{\tau \mathfrak{b} s}^{\dagger}(\rr) T_{\tau}(\rr) \hat{\psi}_{\tau \mathfrak{t} s}(\rr)-\text{H.c.}],
\end{equation}
where $j_{z}$ has a unit of current per area.

We calculate the current density $\langle j_{\alpha}(\rr) \rangle$ in the chiral $d$-wave states using a perturbation theory by expanding the Green's function in a power series of the pair amplitudes $\Delta$. We  retain only the leading-order contributions, and $\langle j_{\alpha}(\rr) \rangle$ is then proportional to $\Delta^* \Delta$, as required by gauge invariance. 
The bulk current distribution calculated in this way for the $\hat{\Gamma}_1$ pair potential is illustrated in Fig.~\ref{Fig:supercurrent}.  Because the current is evaluated for the superconducting state, it is supercurrent without dissipation.  The in-plane currents in bottom and top layers flow respectively  out-of and in-to the centers of AA regions, and nearly compensates each other. This current pattern is closely related to the layer counterflow velocity in the nearly flat moir\'e bands. The total in-plane current $\langle \jj_{\parallel}(\rr) \rangle$ is still finite and circulates around the $\hat{z}$ axis. The out-of-plane current pattern is consistent with the in-plane current flow. Moreover, we numerically find that the current continuity is satisfied,  $\langle j_z(\rr) \rangle =-\boldsymbol{\nabla}\cdot \langle \jj_{\mathfrak{b}}(\rr) \rangle = \boldsymbol{\nabla}\cdot \langle \jj_{\mathfrak{t}}(\rr) \rangle $. Overall, the current flow can be decomposed into two components. One component is the circulation around the $\hat{z}$ axis as shown in Fig.~\ref{Fig:supercurrent}(c), and another component is the current circulation between the two layers. 

The current flow pattern is characterized by two different moments, the magnetic dipole moment $\boldsymbol{m}$ and the magnetic toroidal dipole moment $\boldsymbol{t}$\cite{Spaldin2008}
\begin{equation}
\begin{aligned}
\boldsymbol{m} &= \frac{1}{2}\int d \rr [\rr \times \jj(\rr)],\\
\boldsymbol{t} &= \frac{1}{10} \int d\rr[\rr (\rr\cdot\jj(\rr))-2 r^2 \jj(\rr)].
\end{aligned}
\end{equation}
Because the total bulk current vanishes, both $\boldsymbol{m}$ and $\boldsymbol{t}$ are extensive quantities proportional to the total number of moir\'e unit cells. For the current distribution illustrated in Fig.~\ref{Fig:supercurrent}, $\boldsymbol{m}$ and $\boldsymbol{t}$ per moir\'e unit cell are respectively about $8\times 10^{-3}\mu_B \hat{z}$ and $-2.4d_z\mu_B \hat{z}$, where $\mu_B$ is the Bohr magneton and $d_z$ is the interlayer vertical distance of TBLG.  $\boldsymbol{m}$ and $\boldsymbol{t}$ scale as  $\Delta^* \Delta$, and their exact values therefore depend on the pair amplitudes.  The above numbers should be viewed as an order of magnitude estimation.   The magnetic dipole moment $\boldsymbol{m}$ could be detected by magnetization measurement, while the toroidal dipole moment $\boldsymbol{t}$ could lead to magnetoelectric effect.\cite{Spaldin2008} Polar Kerr effect \cite{Brydon2018} could also be used to probe the time-reversal symmetry breaking in the chiral state. 

\begin{figure*}[t!]
	\includegraphics[width=1.8\columnwidth]{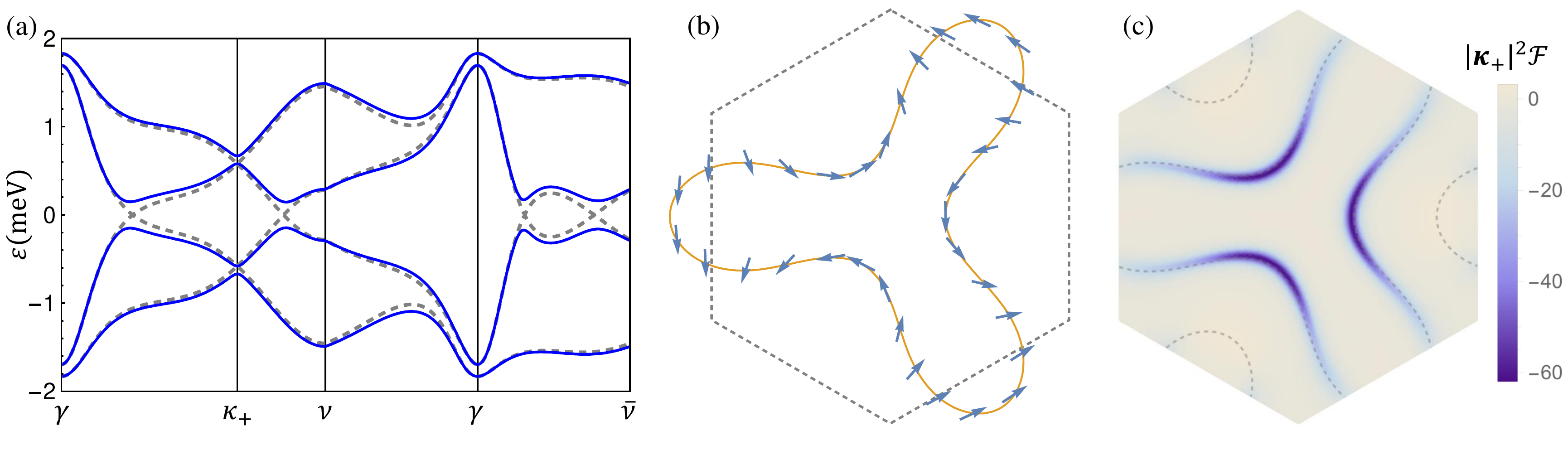}
	\caption{(a) Band structure (blue curves) of $H_{BdG}(\qq)$ determined by using the chemical potential for half-filled lower flat band and the  $\hat{\Gamma}_1$ pair potential. The pair amplitudes are shown in Figs.~\ref{Fig:pair_potential}(c, d) with the overall scale $-4 g_{A_1} \Delta_{\mathfrak{b}}^{(1,+)}(\boldsymbol{0})$ set to be 1 meV. The gray dashed curves show the band structure of $H_{BdG}(\qq)$ without pair potential. (b) The corresponding gap function $\Delta(\qq)$ on the Fermi surface (yellow curve) in the extended moir\'e Brillouin zone. The blue arrows indicate the magnitude and phase of $\Delta(\qq)$. (c) The Berry curvature $\mathcal{F}$ for the two lower bands (blue curves) in (a). The dashed curves show the Fermi surface of the normal state. }
	\label{Fig:chiral}
\end{figure*}

We note that the magnetic field induced by the spontaneous current also scales as  $\Delta^* \Delta$, and therefore, does not enter into the linearized gap equation (\ref{chi}). The feedback effect of the spontaneously generated magnetic field on the pairing order parameter is negligible near the critical temperature.

Our calculation of the supercurrent was initially motivated by the vortex structure in the pair amplitudes. However, even if the vortices in the pair amplitudes are neglected theoretically, for example by taking $\Delta_{\ell}^{(1,-)}$ in $\hat{\Gamma}_1$ to be zero, we find that the bulk circulating supercurrent remains qualitatively the same. The supercurrent is therefore not directed tied to the vortices, although both are  phenomena that arise due to the enlarged moir\'e superlattice.

\section{Topological chiral $d$-wave superconductor}
\label{Sec:TopChiral}
We study the gap structure and topological characters of the $d$-wave superconducting states. The mean-field Hamiltonian for spin-singlet superconductivity can be generally written as
\begin{equation}
\begin{aligned}
H_{MF} &=\int d\rr [\hat{\xi}^\dagger H_{BdG} \hat{\xi} + \hat{\zeta}^\dagger H_{BdG} \hat{\zeta}], \\
\hat{\xi}^\dagger &= (\hat{\psi}_{+ \uparrow}^\dagger, \hat{\psi}_{- \downarrow}) ,
\hat{\zeta}^\dagger = (\hat{\psi}_{+ \downarrow}^\dagger, -\hat{\psi}_{- \uparrow}),\\
\end{aligned}
\end{equation}
where the subscripts $\pm$ of $\hat{\psi}$ again refer to valleys, and $\uparrow$ and $\downarrow$ still label spins, but the layer and sublattice indices are suppressed for conciseness. The  Bogoliubov–--de Gennes (BdG) Hamiltonians $H_{BdG}$ for $\hat{\xi}$ and $\hat{\zeta}$ are the same, which reflects the spin SU(2) symmetry of the spin singlet superconductivity. 

In momentum space, $H_{BdG}$ can be organized as follows
\begin{equation}
H_{BdG}(\qq)=\begin{pmatrix}
H_0(\qq) & \Lambda\\
\Lambda^\dagger & -H_0(\qq)
\end{pmatrix},
\end{equation}
where $\qq$ is the momentum within the moir\'e Brillouin zone. $H_0(\qq)$ is the moir\'e Hamiltonian (including chemical potential term) in valley $+K$, while the moir\'e Hamiltonian in valley $-K$ after performing the particle-hole transformation is given by $-H_0(\qq)$ . The off-diagonal terms $\Lambda$ and $\Lambda^{\dagger}$ are matrix representation of the pair potentials. In our case, $\Lambda$ is independent of $\qq$ because the pairing interaction in (\ref{HBCS}) is local in space.  

It is instructive to project $H_{BdG}(\qq)$ to states on the Fermi surface, leading to the following $2\times 2$ matrix
\begin{equation}
\tilde{H}_{BdG}(\qq)=\begin{pmatrix}
0 & \Delta(\qq)\\
\Delta^*(\qq) & 0
\end{pmatrix},
\end{equation}
where $\Delta(\qq)= \langle u(\qq)|\Lambda| u(\qq) \rangle$, and  $|u(\qq) \rangle$ is a state with momentum $\qq$ on the Fermi surface. $\Delta(\qq)$ is generally referred to as the gap function. If the pair potential is time-reversal symmetric, then $\Lambda=\Lambda^\dagger$ and therefore, $\Delta(\qq)$ is real; otherwise, $\Delta(\qq)$ is generally complex.

We apply the above discussion to chiral $d$-wave states. Fig.~\ref{Fig:chiral}(a) shows the band structure of $H_{BdG}(\qq)$ determined by using the chemical potential for half-filled lower flat band and the  $\hat{\Gamma}_1$ pair potential with pair amplitudes shown in Figs.~\ref{Fig:pair_potential}(c, d); Fig.~\ref{Fig:chiral}(b) illustrates the corresponding gap function $\Delta(\qq)$ on the Fermi surface, which indicates that the chiral $d$-wave state is fully gapped and that the phase of $\Delta(\qq)$ changes by $4\pi$ when $\qq$ moves along the Fermi surface once. These features are generally expected for chiral $d$-wave states.

To characterize the topological property, we calculate the Berry curvature for all occupied bands in Fig.~\ref{Fig:chiral}(a). The corresponding Berry curvature, as shown in Fig.~\ref{Fig:chiral}(c), is strongly peaked near the Fermi surface of the normal state. The Chern number of $H_{BdG}(\qq)$, obtained by integrating the Berry curvature in Fig.~\ref{Fig:chiral}(c), is equal to $-2$; the total Chern number is then $-4$ after taking into account of the spin degeneracy. Table~\ref{Table:Chern} summarizes the dependence of the total Chern number on the chemical potential and the pairing potential, which is reminiscent of the topological character of a model Hamiltonian for chiral $d$-wave superconductivity in monolayer graphene. At $\mu=0$ (Dirac point energy), the energy spectrum of $H_{BdG}(\qq)$ remains {\it gapless} at $\boldsymbol{\kappa}_\ell$ points even if the pair amplitudes in $\hat{\Gamma}_1$ or $\hat{\Gamma}_2$ are finite, which explains the abrupt change in the Chern number as $\mu$ crosses zero.

\begin{table}[b]
	  \caption{ The dependence of the total Chern number on the chemical potential and the pairing potential. $\varepsilon_{\gamma,\pm}$ are the upper ($+$) and lower ($-$) flat band energies at $\gamma$ point. }
	\begin{tabular}{c|c|c}
		\hline
		&  $\varepsilon_{\gamma,-}<\mu<0$ &  $0<\mu<\varepsilon_{\gamma,+}$   \\
		\hline
	$\hat{\Gamma}_1$	& $-4$ & $+4$    \\
	\hline
	$\hat{\Gamma}_2$	& $+4$ & $-4$  \\
	\hline
	\end{tabular}
\label{Table:Chern}
\end{table}

\begin{figure}[t]
	\includegraphics[width=1.0\columnwidth]{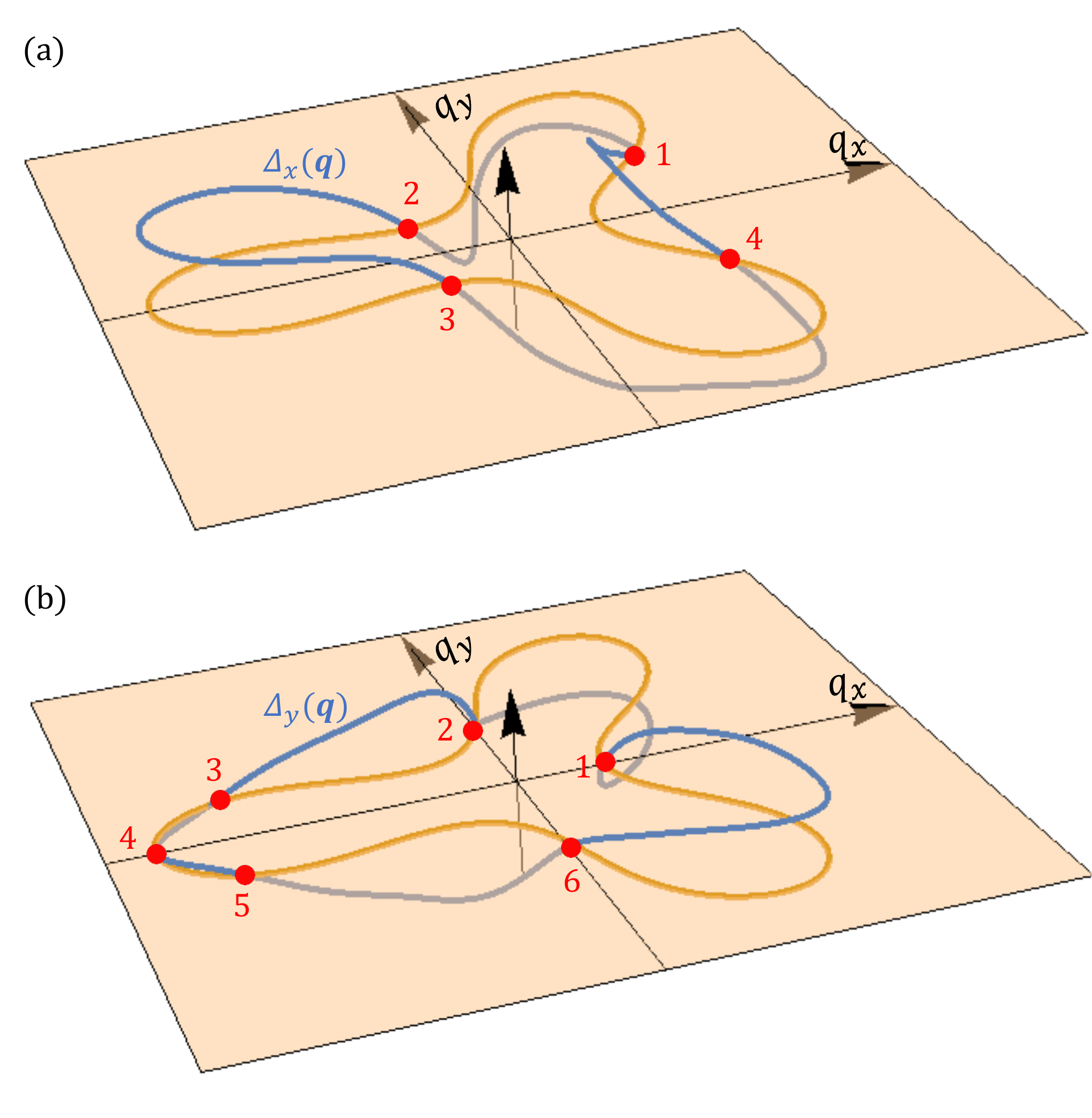}
	\caption{(a) Gap function $\Delta_x(\qq)$ (blue curve) for the nematic pair potential $\hat{\Gamma}_x$ and for $\qq$ on the Fermi surface (yellow curve). $\Delta_x(\qq)$ is real and crosses zeros at the four red points.  (b) Gap function $\Delta_y(\qq)$  for the pair potential $\hat{\Gamma}_y$. $\Delta_y(\qq)$ has six point nodes. The point node $4$ can be annihilated by node $3$ or $5$ when the nematic director $\boldsymbol{\eta}$ is slightly away from $\hat{y}$ axis. Parameters used in the calculation are the same as those in Fig.~\ref{Fig:chiral}.}
	\label{Fig:nematic}
\end{figure}

\section{Nematic $d$-wave superconductor}
\label{Sec:NematicD}
Chiral and nematic $d$-wave pairing order parameters belong to the same multiplet.  Two independent basis functions for nematic $d$-wave pairings are given by
\begin{equation}
\begin{aligned}
\hat{\Gamma}_x & = i(\hat{\Gamma}_1-\hat{\Gamma}_2)/\sqrt{2},\\
\hat{\Gamma}_y & = (\hat{\Gamma}_1+\hat{\Gamma}_2)/\sqrt{2}.
\end{aligned}
\end{equation}
$\hat{\Gamma}_x$ and $\hat{\Gamma}_y$  transform respectively as $d_{x^2-y^2}$ and $d_{xy}$ under $\hat{C}_6$ rotation.
A generic nematic pair potential can be parametrized as
\begin{equation}
\hat{\Gamma}_{\boldsymbol{\eta}}=\eta_x \hat{\Gamma}_x+\eta_y \hat{\Gamma}_y,
\end{equation}
where $\boldsymbol{\eta}=(\eta_x,\eta_y)$ is the nematic director.  A superconductor with the nematic pair potential  preserves time-reversal symmetry but breaks rotational symmetry, which is characterized by a nematic order parameter
\begin{equation}
\boldsymbol{N}=(|\eta_x|^2-|\eta_y|^2, \eta_x^*\eta_y+\eta_y^*\eta_x).
\end{equation}

Because $\hat{\Gamma}_{\boldsymbol{\eta}}$ is time-reversal symmetric, there is no spontaneous bulk supercurrent and the corresponding gap function $\Delta_{\boldsymbol{\eta}}(\qq)$ is real. On the other hand,  $\Delta_{\boldsymbol{\eta}}(\qq)$  integrated over the Fermi surface (FS) vanishes
\begin{equation}
\int_{\qq \in FS} d\qq \Delta_{\boldsymbol{\eta}}(\qq)=0.
\label{FSInt}
\end{equation}
Therefore, $\Delta_{\boldsymbol{\eta}}(\qq)$ must have point nodes on the Fermi surface, as illustrated in Fig.~\ref{Fig:nematic}. $\Delta_x(\qq)$ of the pair potential $\hat{\Gamma}_x$ has four point nodes. However, $\Delta_y(\qq)$ of the pair potential $\hat{\Gamma}_y$ has six point nodes, two of which annihilate each other when $\boldsymbol{\eta}$ deviates slightly away from $\hat{y}$ axis.

Chiral and nematic superconducting order parameters have degenerate superconducting transition temperature $T_c$, as they belong to the same $E_2$ representation. Because of the different gap structure, chiral states are energetically more favored below $T_c$ in a weak-coupling mean-field theory that  takes into account only the superconductivity instability. 

An external uniaxial strain breaks the six-fold rotational symmetry, and therefore, lifts the two-fold degeneracy between $\hat{\Gamma}_x$ and $\hat{\Gamma}_y$. The uniaxial strain tensor $\epsilon_{ij}$ couples linearly to the nematic order parameter $\boldsymbol{N}$, and extrinsically stabilizes nematic superconductivity near $T_c$. An in-plane magnetic field $\boldsymbol{B}_{\parallel}$ also breaks the rotational symmetry, and could play a  role similar to uniaxial strain. An interplay between the $\boldsymbol{B}_{\parallel}$ field and the strain $\epsilon_{ij}$ field can lead to a two-fold anisotropy in the critical in-plane  magnetic field. The nematic superconductivity could also be intrinsically stabilized by density wave fluctuations as proposed by a recent theoretical work\cite{Kozii2018}.

\section{Discussion}
\label{sec:discussion}
In TBLG, optical phonon modes can mediate $d$-wave pairing because of the sublattice pseudospin chirality. It recently became recognized through the study of topological superconductivity that phonon fluctuations  assisted by strong spin-orbital coupling can generate {\it non} $s$-wave pairing interactions.\cite{Fu_Berg,Brydon2014,Kozii2015, Wu2017Nematic}  Our work provides a distinct example in which the sublattice pseudospin chirality enables the $d$-wave pairing without the need of real spin-orbit coupling.
A recent theoretical study showed that acoustic phonons in TBLG can also mediate unconventional pairing such as $d$ wave.\cite{Wu2019rho} 
We note that the pairing mechanism in TBLG is a subject under intense theoretical study, and many different mechanisms are being explored. Our findings on the  properties of $d$-wave states are largely independent of the exact pairing mechanism, as they are mainly controlled by symmetry and topology. We have used a weak-coupling theory, which can be partly justified within our theoretical framework because the pairing energy scale $k_B T_c$ is still an order of magnitude smaller than the bandwidth.  Our results are fully self-consistent within mean field theory, while effects beyond mean field theory, such as fluctuations, are not included in our work.

In summary, we have studied the  pairing order parameters, gap structure and topological character of $d$-wave superconductivity in TBLG. The presence of spontaneous vortices and supercurrent could be a very generic effect for multicomponent superconductivity and superfluidity in superlattices; a theory that incorporates microscopic physics within the superlattice unit cell is crucial to study this effect.  Two-particle bound states with a finite center-of-mass angular momentum were also recently studied in the context of excitons in moir\'e pattern \cite{Wu2018}.

\section{Acknowledgment}
F. W. thanks I. Martin, S. Das Sarma, A. H. MacDonald and P. Jarillo-Herrero for helpful discussions. Work at the University of Maryland was supported
by the Laboratory for Physical Sciences. Work at Argonne National Laboratory was supported by the Department of Energy, Office of Science,
Materials Science and Engineering Division. The author also acknowledges support from a QuantEmX Scientist Exchange Award by ICAM and the Gordon and Betty Moore Foundation for a visit to the University of Michigan, where part of the work was done.

\bibliographystyle{apsrev4-1}
\bibliography{refs}

\end{document}